\documentclass[sigconf, nonacm]{cidr-2027}
\AtBeginDocument{}

\usepackage{mathtools}
\usepackage{tikz}
\usetikzlibrary{positioning,fit,calc,arrows.meta,shapes.geometric,backgrounds,matrix,decorations.pathreplacing}
\usepackage{xcolor}
\usepackage{graphicx}
\usepackage{enumitem}

\usepackage{listings}
\lstdefinestyle{prooftrace}{
    basicstyle=\ttfamily\scriptsize,
    breaklines=true,
    columns=fullflexible,
    keepspaces=true,
    frame=single,
    rulecolor=\color{black!30},
    framerule=0.3pt,
    backgroundcolor=\color{black!2},
    aboveskip=4pt,
    belowskip=4pt,
}
\usepackage{enumitem}

\newcommand{\sys}{EquiTrace}
\newcommand{\sem}[1]{[\![#1]\!]}

\definecolor{stagebg}{HTML}{F0F4F8}
\definecolor{stageedge}{HTML}{4A6480}
\definecolor{stage5bg}{HTML}{E8F0E8}
\definecolor{stage5edge}{HTML}{5A7A4A}
\definecolor{resultedge}{HTML}{B38B3D}
\definecolor{matrixhdr}{HTML}{E0E6EC}
\definecolor{resultbg}{HTML}{FFF5E0}
\definecolor{onmark}{HTML}{2C7A2C}
\definecolor{onmarkfill}{HTML}{D8EBD8}
\definecolor{offmark}{HTML}{8C8C8C}
\definecolor{perpanelbg}{HTML}{F7F9FB}

\begin{document}

\title{Building SQL Equivalence Proofs from Query Rewrite Rules}

\author{Chunyu Chen}
\affiliation{%
  \institution{Simon Fraser University}
  \city{Burnaby}
  \state{BC}
  \country{Canada}
}
\email{chunyu_chen@sfu.ca}
\author{Xinyi Huang}
\affiliation{%
  \institution{Simon Fraser University}
  \city{Burnaby}
  \state{BC}
  \country{Canada}
}
\email{xha88@sfu.ca}

\author{Zhengjie Miao}
\affiliation{%
  \institution{Simon Fraser University}
  \city{Burnaby}
  \state{BC}
  \country{Canada}
}
\email{zhengjie@sfu.ca}

\begin{abstract}

Query rewriting and semantic evaluation of text-to-SQL systems both
depend on deciding whether two SQL queries return the same result on
every database satisfying the schema constraints. Decades of
equivalence-preserving optimizer rules provide a natural basis for
constructive proofs: two queries are equivalent when admitted
rewrites transform them to a common logical expression. Building on
this observation, \sys{} organizes query-rewriting knowledge as a
standalone proof-producing system. It separates rule validation from
proof discovery, explicitly aligns alternative query
representations, and returns the resulting rewrite derivation.
Across three corpus classes, \sys{} proves 648/1,106
expected-equivalent text-to-SQL pairs (58.6\%), 229/266 curated pairs
(86.1\%), and 97/240 LLM-generated rewrites whose results agree under
benchmark execution (40.4\%). In a pinned comparison on the query rewrite
corpus, \sys{} proves 97 pairs, compared with 35 for QED and 38 for
SQLSolver. An analysis of 1,242 recorded proofs shows that coverage
depends on a small number of transformations,
while many frequently used rules are individually substitutable.
The resulting derivations make each proof replayable and expose the
specific transformations on which it depends.

\end{abstract}

\maketitle

\section{Introduction}\label{sec:intro}

Query equivalence is foundational in database systems, with wide applications in query optimization~\cite{chaudhuri98optimization}. Recent advances in large language models (LLMs) have renewed interest in determining whether syntactically different queries produced by query-rewriting and text-to-SQL systems return the same result on every database satisfying the schema constraints.



A long line of work approaches equivalence checking through
symbolic reasoning: SPES~\cite{spes},
SQLSolver~\cite{sqlsolver}, QED~\cite{qed}, and
VeriEQL~\cite{verieql} translate SQL into algebraic or
SMT-friendly representations and reason about their
equivalence. While principled, supporting the full
diversity of SQL requires ever-richer models of bag
semantics, NULLs, subqueries, and
integrity constraints --- leading to unsatisfying empirical
coverage of practical queries, in the literature and in our
experiments (Section~\ref{sec:coverage}). That raises a
complementary question: must every query pair be proved by
reconstructing its full semantics inside a solver, or can
we reuse the equivalence knowledge database people have
accumulated over decades?

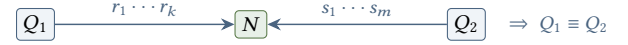
\begin{figure}[t]
  \centering



\begin{tikzpicture}[
    font=\small\sffamily,
    >={Stealth[length=1.8mm,width=1.6mm]},
    box/.style={
        draw=stageedge, fill=stagebg, rounded corners=1.5pt,
        line width=0.4pt, inner sep=2.5pt, align=center,
    },
    goal/.style={box, draw=stage5edge, fill=stage5bg},
    lbl/.style={font=\small\sffamily\itshape, text=stageedge, anchor=west},
    arr/.style={->, line width=0.55pt, draw=stageedge},
    rlbl/.style={font=\footnotesize\sffamily, text=stageedge},
]
\node[lbl] at (-0.2, 0.82) {(a) Query optimization: pick the most cost-efficient member of the class};
\node[box] (q)  at (0.35, 0.22) {$Q$};
\node[box] (es) at (2.85, 0.22) {$\{E_1, E_2, \ldots\}$};
\node[goal] (pl) at (6.15, 0.22) {cheapest plan};
\draw[arr] (q) -- node[above, rlbl] {rewrite rules} (es);
\draw[arr] (es) -- node[above, rlbl] {cost model} (pl);

\node[lbl] at (-0.2, -0.50) {(b) Equivalence proving: connect two queries inside the class};
\node[box] (q1) at (0.45, -1.08) {$Q_1$};
\node[goal] (n) at (3.30, -1.08) {$N$};
\node[box] (q2) at (6.15, -1.08) {$Q_2$};
\draw[arr] (q1) -- node[above, rlbl] {$r_1 \cdots r_k$} (n);
\draw[arr] (q2) -- node[above, rlbl] {$s_1 \cdots s_m$} (n);
\node[rlbl, anchor=west] at (6.60, -1.08)
    {$\Rightarrow\; Q_1 \equiv Q_2$};
\end{tikzpicture}
  \vspace{-3mm}
  \caption{The same rewrite vocabulary supports two objectives:
    optimization explores an equivalence class to select an efficient
    plan, whereas equivalence proving constructs derivations from
    $Q_1$ and $Q_2$ to a common expression $N$.}
  \label{fig:concept}
  \vspace{-3mm}
\end{figure}

Query optimizers already encode a substantial body of equivalence
knowledge. A logical transformation rule states a reusable semantic
identity: whenever its applicability conditions hold, replacing one
logical expression with another preserves the query result.
Optimization composes such transformations to explore alternative
plans and select an efficient one. The same transformations can instead
serve as proof steps. If admitted rewrites transform both $Q_1$ and
$Q_2$ to a common logical expression $N$, the two resulting
derivations constitute a constructive proof that
$Q_1 \equiv Q_2$ (Figure~\ref{fig:concept}).

Narasayya and Chaudhuri recently explored this observation in QO-Verify~\cite{qoverify}: rather than building a separate semantic solver, it extends the SQL Server optimizer and searches the logical-expression spaces generated for the two queries; a common expression establishes equivalence, because each optimizer transformation is assumed to preserve semantics. A production optimizer, their results show, already contains enough equivalence knowledge to verify a substantial fraction of LLM-generated rewrites.

This raises a broader question: \textbf{can query-rewriting
knowledge form the basis of a general, standalone SQL
equivalence proof system?} Existing optimizers provide useful infrastructure for equivalence checking, including internal query representations, transformation libraries, and search strategies. However, our experience adapting Calcite for equivalence proving reveals three mismatches between optimization and proof construction. First, useful equalities may be oriented in only one direction; for example, Calcite includes guarded join-to-semi-join rules, but not the uniqueness-guarded reverse transformations required by some proofs. Second, some equivalence proofs require transformations that operate beyond a local rewrite. Pairwise join rules can enumerate alternative join trees, whereas canonicalization may require inspecting and rebuilding an entire commutative join region. Third, equivalent expressions may remain syntactically distinct, such as \texttt{a=b} and \texttt{b=a} or differently ordered conjunctions, and therefore require explicit normalization before their derivations can converge.




A standalone prover can combine inference rules from optimizer transformations, verified rewrite libraries, integrity-constraint reasoning, and semantic canonicalizations whose purpose is to expose equivalence rather than produce an efficient plan. Unlike a prover embedded in a proprietary optimizer, such a system can also expose the derivation, making explicit both the transformations that justify the result and the assumptions on which it depends.



We present \textbf{\sys{}}, a standalone rewrite-based SQL
equivalence prover built on three principles: it
\emph{separates rule validity from proof discovery}
(transformation rules are inputs, assumed sound; proving
searches for their compositions); it makes
\emph{representation alignment} first-class, bridging
structurally different encodings of the same query; and it
treats the \emph{derivation as the output} --- every proof
records a replayable sequence of transformations reaching a
common logical expression. The implementation uses Apache
Calcite~\cite{calcite}, but the principles do not depend on
Calcite or on any single optimizer's inventory; users may
supply their own rule sets.

Our evaluation measures the coverage of this proof model and identifies its remaining limitations. \sys{} proves 648/1\,106
expected-equivalent text-to-SQL-derived pairs (58.6\%), 229/266
curated pairs (86.1\%), including 210/232 Calcite rule tests, and 97/240 LLM-generated rewrites whose queries agree under
benchmark execution (40.4\%). On the last corpus, a common subset yields 97 proofs for \sys{}, 35 for QED, and 38 for SQLSolver, with 114 pairs proved by their union. Analysis shows that the remaining LLM-rewrite failures arise either from insufficient search capacity in large join regions or from missing representations, particularly for union branches, outer joins, and window functions.

\noindent\textbf{Contributions.} Our contributions are summarized below:
\begin{itemize}[leftmargin=*]
\item We propose a proof model for SQL equivalence based on compositions of
independently validated rewrite rules, with the resulting derivation
as a first-class artifact
(Section~\ref{sec:background}--Section~\ref{sec:system}).

\item We develop \sys{}, a standalone and extensible proof-building framework that explicitly
separates rule validation, representation alignment, and proof search,
and serializes successful rule applications as replayable certificates
(Section~\ref{sec:system}).

\item We evaluate \sys{} across three workload classes, including a
comparison with prior formal verification-based provers with analyses of both transformation-level necessity and the remaining proof gaps
(Section~\ref{sec:coverage}--Section~\ref{sec:limits}).
\end{itemize}


\section{Background and Related Work}\label{sec:background}

\subsection{Query rewriting and logical equivalence}\label{sec:rewriting}

Database optimizers represent queries as logical expressions and apply \emph{semantics-preserving rewrite rules}. A rule $r$ rewrites  an expression $E$ into $E' = r(E)$ such that $\sem{E} =
\sem{E'}$ whenever its side condition $\phi_r$ holds.
Examples include predicate pushdown, decorrelation, and constraint-dependent rewrites whose side conditions rely on keys, foreign keys, or nullability. Optimizers compose these transformations, organize equivalent expressions in a memo, and search for the lowest-cost expression equivalent to an input $Q$.


Equivalence proving can be viewed as a complementary use of the same transformation framework. Rather than searching for the lowest-cost representative of an equivalence class, a verifier determines whether two input queries, $Q_1$ and $Q_2$, can be connected by equivalence-preserving transformations as in Figure~\ref{fig:concept}.

The resulting rewrite derivations constitute a checkable proof: each step identifies a equivalence-preserving rule and establishes that its side conditions hold.


\subsection{Existing equivalence provers}\label{sec:provers}

The major existing family translates query pairs into formal semantic representations and discharges equivalence using a solver or proof system. This includes semiring-based encodings with automated reasoning (UDP~\cite{udp}, SPES~\cite{spes}, SQLSolver~\cite{sqlsolver}, and QED~\cite{qed}), bounded verification (VeriEQL~\cite{verieql}), and proof-assistant-based approaches (EQUITAS~\cite{equitas} and HoTTSQL~\cite{hottsql}). These methods provide principled guarantees but require the relevant SQL semantics to be encoded within the underlying formal system. Rewrite-based verification instead treats previously validated transformations as reusable lemmas. WeTune~\cite{wetune} discovers candidate rewrites and verifies them offline using SMT, whereas LLM-based systems such as SPA~\cite{huang2026spa}, GenRewrite~\cite{genrewrite} and Lithe~\cite{lithe} motivate the rewriter-output workload evaluated in \S\ref{sec:coverage}.


\subsection{QO-Verify: the optimizer as verifier}\label{sec:qoverify}

Narasayya and Chaudhuri~\cite{qoverify} show that two queries can be verified as equivalent when their optimizer search spaces contain a common logical expression. QO-Verify implements this idea within the  SQL Server optimizer and verifies approximately 37\% of the LLM-generated rewrites in its evaluation, demonstrating the value of existing optimizer transformations for equivalence reasoning.



We take a complementary approach by using query-rewriting knowledge to construct a \emph{standalone} proof system rather than treating the optimizer itself as the verifier.
Each rewrite is an inference rule, and two rewrite sequences that reach a common expression constitute an explicit equivalence proof. This design makes the rule library 
and proof search independently extensible, while enabling proofs to be inspected and replayed and allowing each rule to be associated with an explicit trust status. Moreover, failed proofs can be analyzed in terms of missing transformations. 



\section{\sys}\label{sec:system}

\subsection{A proof-oriented rewrite system}
\label{sec:rewrite-system}

\sys{} organizes heterogeneous query-rewriting knowledge as an
executable proof-oriented rewrite system. Its \emph{primitive proof
rules} are executable, semantics-preserving transformations with
explicit applicability conditions. They include optimizer rules,
proof-oriented inverse rewrites, and schema-dependent equalities.
A \emph{proof strategy} is a deterministic, bounded, and possibly
global policy for applying such transformations, for example to
align alternative encodings of existence or to select a canonical
representative for a commutative join region. 
\sys{} combines admitted proof rules and strategies with exploratory search
and certificate construction.

The individual equalities are largely established results rather
than new identities. Our contribution is to make their admission,
orientation, composition, provenance, and application explicit in a
standalone proof-producing system. Optimizer transformations provide
an important subset of the rule vocabulary, while additional rules
and strategies support representation alignment,
integrity-constraint reasoning, and proof-oriented directions absent
from the selected optimizer configuration.


\subsection{Design principles}\label{sec:principles}

\noindent\textbf{P1: Separate rule validation from proof discovery.}
Each transformation $r$ admitted by \sys{} is assumed to preserve
semantics whenever its applicability condition $\phi_r$ holds:
\[
  \phi_r(E) \Rightarrow \sem{E} = \sem{r(E)}.
\]
Establishing this rule-level obligation is orthogonal to query-level
proof search. It may rely on formal verification, externally verified
rewrite libraries, upstream validation, or rule-specific testing.
\sys{} does not re-prove the transformation for every query pair;
instead, it composes admitted transformations into a derivation.
Rule validation can therefore be performed once and reused across
many proofs.

\noindent\textbf{P2: Treat cross-query representation alignment as an
explicit proof objective.}
Optimizers routinely normalize logical expressions to support
planning, but their objective is to find an efficient plan for one
query, not to make two independently written queries converge to the
same representation. For equivalence proving, a transformation may
therefore be valuable solely because it reduces representational
distance between the two inputs. \sys{} separates deterministic
alignment from exploratory rewrite search, records alignment steps as
part of the derivation, and evaluates their contribution
independently.

\noindent\textbf{P3: Treat the derivation as the result.}
A positive verdict should expose why the queries are equivalent, not
merely report that a common expression was found. \sys{} therefore
returns an explicit rewrite proof containing the transformations used,
their relevant side conditions, and the common witness, which can be serialized as a replayable certificate. This design
supports human inspection, mechanical replay, and explicit accounting
of the assumptions on which the result depends.

\subsection{Proof model}\label{sec:proof-model}

Let $\Gamma$ denote the proof environment, including the schema,
integrity constraints, and set of transformations. Each
transformation admitted under $\Gamma$ is executable and replayable,
has explicit applicability conditions, and is required to preserve semantics whenever those conditions hold.
In the following, $Q_1$ and $Q_2$ denote the normalized logical plans obtained from the two input SQL queries.


We write $\Gamma \vdash Q \xRightarrow{\pi} E$
when $\pi$ is a sequence of transformations admitted under $\Gamma$
that transforms logical expression $Q$ into $E$. Given inputs $Q_1$
and $Q_2$, \sys{} searches for
\[
  \Gamma \vdash
  Q_1 \xRightarrow{\pi_1} E_1
  \qquad\text{and}\qquad
  \Gamma \vdash
  Q_2 \xRightarrow{\pi_2} E_2
\]
such that
\[
  E_1 \equiv_{\mathrm{wit}} E_2,
\]
where $\equiv_{\mathrm{wit}}$ requires the expressions to represent
the same logical plan with compatible output contracts.

A successful proof is
\[
  \pi = (\Gamma,Q_1,Q_2,\pi_1,\pi_2,E),
\]
where $E$ is the common witness represented by $E_1$ and $E_2$.
Each branch records the transformations, their input and output expressions, and any schema or semantic facts on which their applicability depends. If no witness is discovered within the configured search budget,
\sys{} returns \texttt{UNKNOWN}; failure to find a derivation is not
evidence of inequivalence. 

\begin{figure*}[t]
\centering

\resizebox{\textwidth}{!}{%
\begin{tikzpicture}[
  >=Latex,
  panel/.style={
    draw,
    rounded corners=2pt,
    inner sep=3mm,
    minimum height=\PanelHeight,
    anchor=north west
  },
  plan/.style={
    draw,
    rounded corners=2pt,
    align=center,
    inner sep=2mm,
    font=\scriptsize,
  },
  witness/.style={
    draw,
    rounded corners=2pt,
    align=center,
    inner sep=2.5mm,
    font=\scriptsize,
    thick
  },
  rule/.style={
    font=\scriptsize,
    align=center,
    inner sep=1pt
  },
  every path/.style={thick}
]

\def\PanelHeight{5.9cm}

\node[panel, draw=stageedge, fill=stagebg] (panel-a) at (0,0) {%
\begin{minipage}[t][][t]{0.24\textwidth}
\textbf{(a) Queries and assumption}




\smallskip
{\scriptsize
\textbf{Schema:}
\texttt{R(id, name)},
\texttt{S(rid, rname, sname)}
\smallskip
\textbf{Constraint:}
\(\operatorname{UNIQUE}(S.\mathit{rid})\)
}


\textbf{\(Q_1\): existence form}

\medskip

\colorbox{gray!10}{%
\parbox{\dimexpr\linewidth-2\fboxsep\relax}{%
\scriptsize\ttfamily
SELECT r.id FROM R AS r\\
WHERE EXISTS (\\
\quad SELECT 1 FROM S AS s\\
\quad WHERE s.rid = r.id);\\
}}

\textbf{\(Q_2\): join form}

\colorbox{gray!10}{%
\parbox{\dimexpr\linewidth-2\fboxsep\relax}{%
\scriptsize\ttfamily
SELECT r.id FROM S AS s\\
JOIN R AS r\\
\quad ON s.rid = r.id;
}}


\vfill
\vspace{2mm}

{\scriptsize
\setlength{\baselineskip}{2.5mm}
\noindent
Without uniqueness, \(Q_2\) may duplicate an \(R\) row when
several \(S\) rows match, whereas \(Q_1\) returns it once.
\par
}
\end{minipage}
};

\node[
  panel,
  draw=stage5edge,
  fill=stage5bg,
  minimum width=9.2cm
] (panel-b) at ($(panel-a.north east)+(5mm,0)$) {};

\node[
  font=\bfseries,
  anchor=north
] (panel-b-title)
  at ($(panel-b.north)+(0,-4mm)$)
  {(b) Two derivations to a common witness};

\node[plan] (q1)
  at ($(panel-b.north west)+(23mm,-15mm)$)
  {\(\displaystyle Q_1\)\\
   {\scriptsize correlated \texttt{EXISTS}}};

\node[plan] (q2)
  at ($(panel-b.north east)+(-23mm,-15mm)$)
  {\(\displaystyle Q_2\)\\
   {\scriptsize explicit inner join}};

\node[plan] (semi)
  at ($(q1)+(0,-14mm)$)
  {\(\displaystyle
     \pi_{r.id}
     \bigl(
       R \ltimes_{r.id=s.rid} S
     \bigr)\)};

\node[witness] (common)
  at ($(panel-b.north)+(0,-50mm)$)
  {\(\displaystyle
     E =
     \pi_{r.id}
     \bigl(
       R \bowtie_{r.id=s.rid} S
     \bigr)\)\\
   {\scriptsize common logical expression}};

\draw[->]
  (q1) --
  node[rule, right] (lab-exists)
    {\textsc{ExistsLowering}}
  (semi);

\draw[->]
  (semi) --
  node[
    rule,
    left,
    text width=2.85cm
  ] (lab-semi)
    {\textsc{SemiJoinUniqueRightToInner}\\
     requires \(\operatorname{UNIQUE}(S.\mathit{rid})\)}
  (common);

\draw[->]
  (q2) --
  node[
    rule,
    right,
    text width=2.6cm
  ] (lab-join)
    {\textsc{JoinCommute}\\
     \textsc{NormalizeJoinCondition}}
  (common);

\node[panel, draw=resultedge, fill=resultbg] (panel-c)
  at ($(panel-b.north east)+(5mm,0)$) {%
\begin{minipage}[t][][t]{0.24\textwidth}
\textbf{(c) Executable certificate}

\vspace{2mm}

{\scriptsize
\textbf{Assumption}

\begin{quote}
\(\operatorname{UNIQUE}(S.\mathit{rid})\)
\end{quote}

\textbf{Left derivation}


\begin{enumerate}[
  label=(\arabic*),
  labelindent=0pt,
  leftmargin=3em,
  labelsep=0.3em,
  itemsep=1pt,
  topsep=1pt,
  parsep=0pt
]

  \item
    \texttt{\detokenize{exists_subquery_lowering}}
  \item
    \texttt{\detokenize{semi_join_unique_right_to_inner}}
    
    \emph{Evidence:}
    \(\operatorname{UNIQUE}(S.\mathit{rid})\)
\end{enumerate}

\textbf{Right derivation}

\begin{enumerate}[
  label=(\arabic*),
  labelindent=0pt,
  leftmargin=3em,
  labelsep=0.3em,
  itemsep=1pt,
  topsep=1pt,
  parsep=0pt
]

  \item
    \texttt{\detokenize{join_commute}}

  \item
    \texttt{\detokenize{normalize_join_condition}}
\end{enumerate}

\textbf{Witness}

\[
  \pi_{r.id}
  \bigl(
    R \bowtie_{r.id=s.rid} S
  \bigr)
\]
\textbf{Replay}
\[
  \boxed{\texttt{VALID\_EXACT}}
\]

\vspace{1mm}
{\scriptsize
\setlength{\baselineskip}{2.5mm}
\noindent
Changing or removing the uniqueness constraint invalidates replay.
\par
}
}
\end{minipage}
};

\end{tikzpicture}%
}
\vspace{-6mm}
\caption{
A simplified \sys{} proof.
Query \(Q_1\) expresses existence with a correlated subquery,
whereas \(Q_2\) uses an inner join.
Under bag semantics, their equivalence depends on the uniqueness of
\(S.\mathit{rid}\).
The left derivation lowers \texttt{EXISTS} to a semi-join and then
uses the recorded uniqueness constraint to obtain an inner join.
The right derivation commutes and normalizes the explicit join.
Both branches reach the common expression \(E\).
}
\label{fig:proof-example}
\vspace{-3mm}
\end{figure*}
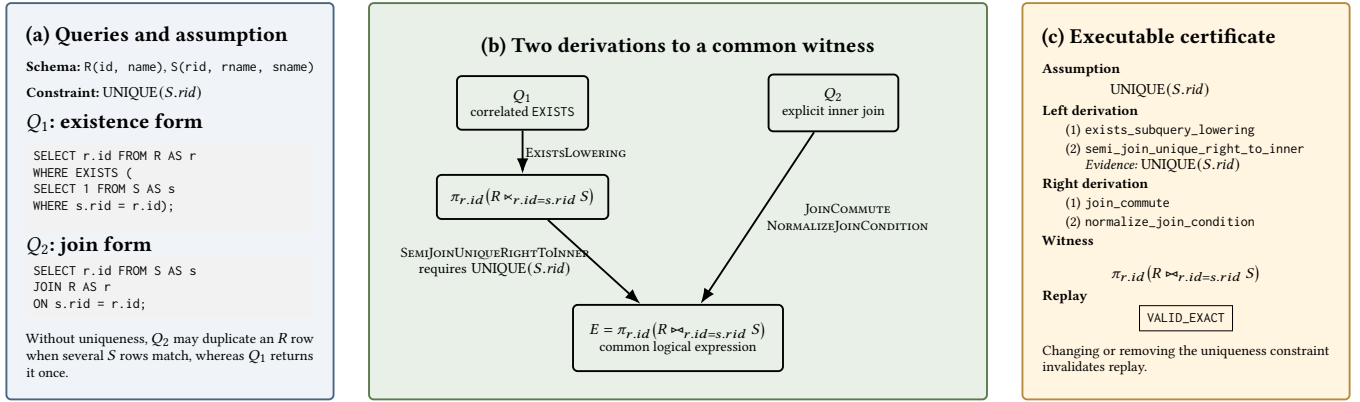

\noindent\textbf{A running example.} Figure~\ref{fig:proof-example} illustrates the proof object
constructed by \sys{} on a minimal pair. Query $Q_1$ expresses
existence through a correlated \texttt{EXISTS}, whereas $Q_2$ uses
an inner join. Under bag semantics, their equivalence depends on
the declared uniqueness of $S.\mathit{rid}$: without it, the join
may duplicate an $R$ row that the existence query returns only once.
Rather than transforming one query directly into the other, \sys{}
constructs two derivations to a common logical expression $E$.
On the left branch, \sys{} lowers \texttt{EXISTS} to a semi-join
and applies \textsc{SemiJoinUniqueRightToInner}. This proof-oriented
rule is the reverse of the semi-join direction provided by the
selected Calcite rule library, and its applicability depends on the
recorded uniqueness constraint. On the right branch, join
commutation and predicate normalization produce the same logical
expression. The shared memo detects this common expression and
establishes convergence of the two derivations.
If the uniqueness declaration is removed, the conversion is not
applicable and \sys{} returns \texttt{UNKNOWN}. Replay succeeds
under the recorded schema and rejects the certificate when the
required constraint is removed. The certificate records both
branches, their before-and-after expressions, the relevant side
condition, and the common witness.

\subsection{Architecture}\label{sec:arch}

\begin{figure}[t]
  \centering
  \includegraphics[width=0.9\linewidth]{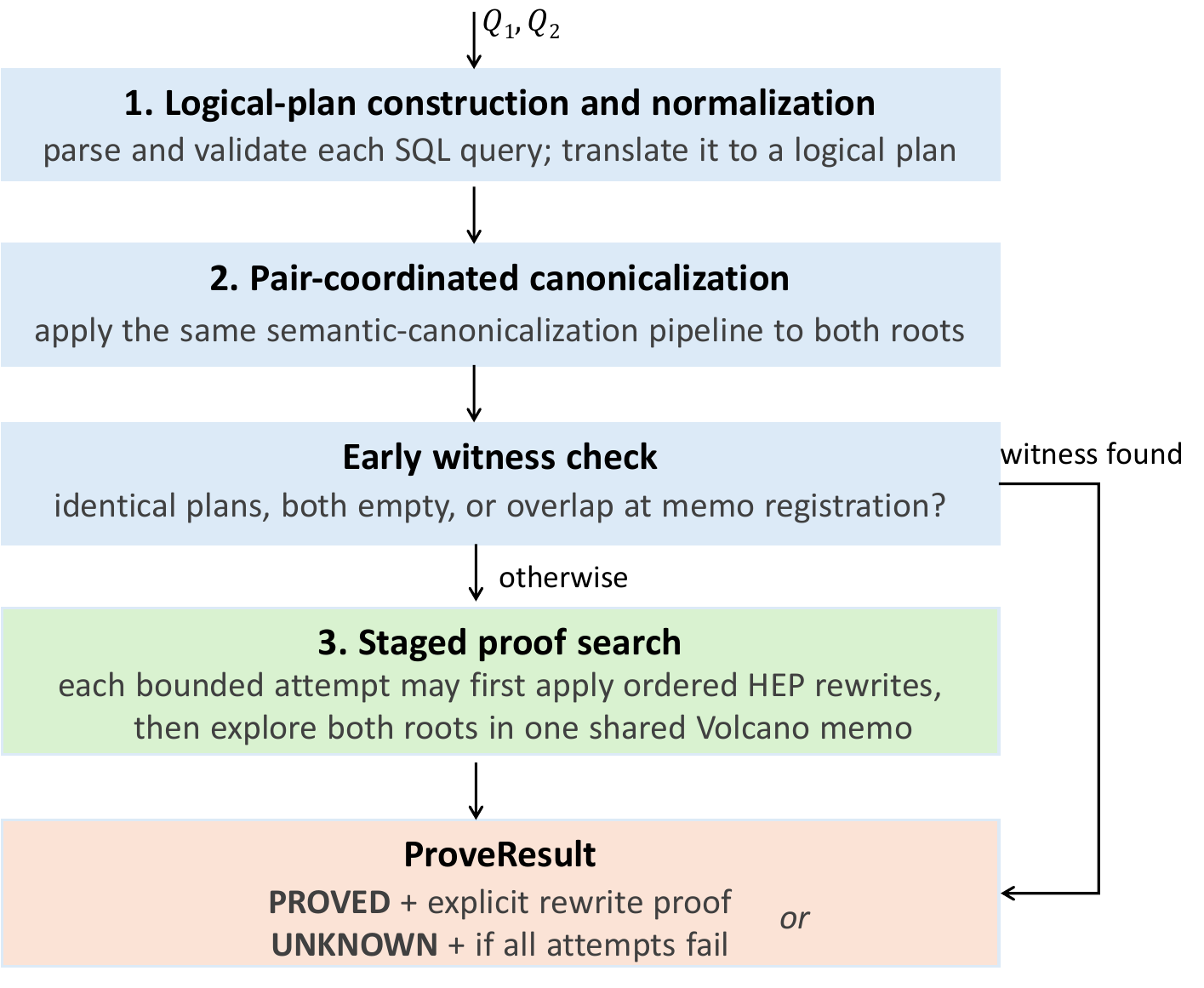}
  \vspace{-3mm}
  \caption{\sys{}'s proof pipeline. }
\label{fig:pipeline}
  \vspace{-6mm}
\end{figure}

\sys{} constructs the two branches of a proof in three stages,
illustrated in Figure~\ref{fig:pipeline}.
\sys{} separates deterministic rewriting from exploratory search
because they serve different proof roles and use different execution
mechanisms. 
A Calcite-based front end produces the normalized logical plans.
Pair-coordinated alignment and preparation
then reduce representational differences under bounded,
deterministic policies. Only the final search phase explores
alternative expressions in a shared memo. Figure~\ref{fig:pipeline}
shows this division of labor.

\noindent\textbf{Logical-plan construction and normalization.}
Each SQL query is parsed and validated against the schema,
canonicalized at the SQL level, and translated by Calcite into a
logical \texttt{RelNode} tree. A mode-specific normalization pipeline
then applies ordered \texttt{HepPlanner} programs and imperative
transformations, including subquery removal and decorrelation. The
resulting logical plans are the inputs to \sys{}'s alignment and
proof-search stages.

\noindent\textbf{Pair-coordinated alignment.}
Starting from the two normalized logical plans, \sys{} applies a
fixed pipeline of semantic canonicalizers to both roots. 
Each pass runs every enabled canonicalizer on both sides; passes repeat until both roots are stable or a fixed pass limit is reached. 
These transformations rewrite alternative encodings
toward shared canonical forms before exploratory search. Alignment
does not itself establish equivalence: it only changes the
representations from which a witness may subsequently be detected.
Every successful application is recorded with its side, pass number,
input and output fingerprints, and resulting logical plan.

\noindent\textbf{Early witness checks.}
After alignment, \sys{} checks whether the two plans
already establish a common witness. This occurs when their plans are identical, when both are proved empty, or when their roots enter
the same equivalence class upon memo registration. Such cases require
no exploratory rule firing. Their certificates record the alignment
steps, if any, and the resulting witness; cases that do not close at
this point proceed to preparation and memo search.


\noindent\textbf{Staged proof search.}
Proofs that do not close after canonicalization proceed through a
sequence of bounded search attempts. Each attempt may first apply an
ordered set of semantics-preserving rewrites to expose operator
structures useful to that attempt. These rewrites are normally
applied to both plans. The resulting plans are then registered in a shared Volcano memo, where the attempt's rule set explores alternative equivalent expressions under a fixed search budget.
If an attempt does not find a common expression, the next may use
additional rewrites, a broader rule set, or a larger budget. One
configuration may rewrite only one side before retrying; the others
advance both sides together. Exhausting all configured attempts
yields \texttt{UNKNOWN}. 

\noindent\textbf{Proof construction.}
As canonicalization and search proceed, \sys{} records the
transformations applied to each input. Proofs are successfully found through memo search when the two branches reach the same
equivalence class. \sys{} packages the two derivations and their
common logical expression as a replayable certificate. Replay
re-executes the recorded transformations and accepts the certificate
only if both branches reach the same witness.


\noindent\textbf{Calcite instantiation.}
Our prototype represents logical plans as Calcite
\texttt{RelNode} trees. \texttt{HepPlanner} executes initial normalization, pair-coordinated canonicalization, and the bounded
rewrites used before individual search attempts; exploratory search
registers both roots in one shared \texttt{VolcanoPlanner} memo.
We use Calcite rules when they express the required transformation
and add custom transformations only for proof-oriented directions,
schema-dependent conditions, or whole-plan canonicalizations absent
from the selected Calcite configuration.

\subsection{Proof replay}
\label{sec:trace}

\sys{} records each successful proof for mechanical replay. The replay
verifier re-executes the named transformations with their original implementations and accepts only if both branches reproduce the
recorded common witness. Replay therefore checks that the recorded
derivation is reproducible; rule-level semantic validity is the
separate obligation described in \S\ref{sec:principles}.
Each recorded transformation is referenced by a stable identifier, which replay resolves through the same rule used during proof discovery.

\section{How Far Can Rewrite-Based Proving Go?}\label{sec:coverage}

\subsection{Experiment setup}\label{sec:workloads}

We evaluate \sys{} on 3,233 query pairs from three corpus classes
(Table~\ref{tab:passrates}).
\begin{itemize}[leftmargin=*]
\item 
The \emph{Text-to-SQL-derived} class contains 2,607 pairs from the
BIRD and LeetCode query corpora, as collected by prior
work~\cite{parseval,verieql}. These pairs were
constructed independently of \sys{} and without knowledge of its
transformation library.
\item The \emph{curated equivalence} class contains 295 pairs: 63 collected
from the literature~\cite{verieql} and 232 from Calcite's
optimizer-rule test suite. The former consists of manually curated equivalence cases, while the latter evaluates whether the prover can recover Calcite’s rule-level transformations. 
\item The \emph{LLM-generated rewrite} class contains 331 candidate
rewrites over 167 distinct original queries produced by two LLMs: GPT-5.4 and
SPA-32B-SI, a specialist model fine-tuned from
Qwen3-32B~\cite{huang2026spa}. The source queries come from DSB,
TPC-H, TPC-DS, and StackOverflow workloads.
Table~\ref{tab:rewriter-corpus} reports the corpus composition.
The originals are structurally complex: they reference a
median of six base tables, 78\% reference at least five, and the
maximum is 32. They contain a median of 18 relational operators
(\(p90=48\), maximum 89).
\end{itemize}

\begin{table}[t]
\centering
\footnotesize
\caption{Composition of the LLM-generated rewrite corpus.}
\vspace{-3mm}
\label{tab:rewriter-corpus}
\begin{tabular}{lrr}
\toprule
Source & Rewrite pairs & Original queries \\
\midrule
TPC-DS        & 178 & 90 \\
DSB           &  77 & 39 \\
TPC-H         &  44 & 22 \\
StackOverflow &  32 & 16 \\
\midrule
Total         & 331 & 167 \\
\bottomrule
\end{tabular}
\vspace{-2mm}
\end{table}

\noindent\textbf{Labels.}
The equivalence labels of pairs in curated suites are known in prior work. The text-to-SQL-derived and LLM-generated corpora use
execution-based agreement labels. For BIRD and LeetCode, we use the
execution results supplied by prior work~\cite{parseval}. For the
LLM-generated rewrites, we execute each original query and candidate
rewrite using the testing tool provided in the corresponding benchmark to generate test databases at the 10GB scale. These labels establish agreement on the tested instances, not
semantic equivalence over all databases satisfying the schema.

\noindent\textbf{Metrics.} 
Our primary metric is proof rate over pairs labeled or constructed
as equivalent:
\[
  \textsc{ProofRate}
  =
  \frac{\text{expected-equivalent pairs proved by \sys{}}}
       {\text{expected-equivalent pairs}}.
\]
Parse failures, crashes, timeouts, and
\texttt{UNKNOWN} outcomes count as misses. 

\noindent\textbf{System configuration.}
The evaluated \sys{} configuration contains 15 semantic canonicalization
components, six bounded HEP rewrite programs, and 83 memo-search rules. It is built on Apache Calcite 1.38.0. 
\sys{} does not treat Calcite's complete optimizer rule set as a
proof library. We admit only logical transformations intended to
preserve query results and whose applicability conditions can be
checked from the logical plan and supplied schema. We exclude
physical-planning, costing, materialization, and planner-internal
rules, as well as transformations that fail our preliminary
system-wide validation (e.g., Calcite's \textsc{SemiJoinRemove} Rule). The resulting rule set is fixed
before evaluation and applied uniformly to every input; rules are
never enabled or disabled for individual benchmark cases.

\noindent\textbf{Baselines.}  We compare \sys{} with the executable equivalence provers QED and SQLSolver on the 331 LLM-generated rewrite pairs.
All three systems receive the same SQL pairs and schema constraints;
baseline versions, timeout settings, and execution environments are
reported in Section~\ref{sec:xprover}


\begin{table}[t]
  \centering
  \footnotesize
  \caption{Coverage across the three benchmark classes.
    Exp.EQ $=$ expected-equivalent cases; Prf \% $=$
    proved~/~Exp.EQ (\S\ref{sec:workloads}).}
  \label{tab:passrates}
  \vspace{-3mm}
  \begin{tabular}{llrrrr}
    \toprule
    Class & Suite & Cases & Exp.EQ & Proved & Prf\,\% \\
    \midrule
    Text-to-SQL- & BIRD            & 1\,432 & 462 & 404 & 87.4 \\
    derived & LeetCode & 1\,175 & 644 & 244 & 37.9 \\
    \cmidrule{2-6}
           & subtotal & \textbf{2\,607} & \textbf{1\,106} & \textbf{648} & \textbf{58.6} \\
    \midrule
    Curated & Literature & 63 & 34 & 19 & 55.9 \\
    suites & Calcite test suite & 232 & 232 & 210 & 90.5 \\
    \cmidrule{2-6}
           & subtotal & \textbf{295} & \textbf{266} & \textbf{229} & \textbf{86.1} \\
    \midrule
    LLM-gen. & GPT-5.4      & 167 & 113 & 33 & 29.2 \\
    rewrites  & SPA-32B-SI   & 164 & 127 & 64 & 50.4 \\
    \bottomrule
  \end{tabular}
  \vspace{-3mm}
\end{table}

\subsection{Coverage by corpus class}\label{sec:realwl}

Table~\ref{tab:passrates} reports coverage across the three corpus
classes. On the text-to-SQL-derived corpora, \sys{} proves 648 of
1\,106 expected-equivalent pairs (\textbf{58.6\%}). Coverage is
substantially higher on BIRD than on LeetCode: 404/462
(\textbf{87.4\%}) versus 244/644 (\textbf{37.9\%}). 

On the curated suites, \sys{} proves 229 of 266
expected-equivalent pairs (\textbf{86.1\%}), including 210/232
(\textbf{90.5\%}) from Calcite's optimizer rule tests. This suite is
particularly well matched to a rewrite-based prover: each pair is
derived from specific Calcite transformations, and \sys{} can reuse
an admitted implementation of that transformation rather than
reconstructing the equality in a separate semantic formalism. For
context, prior work reports 95/232 for SPES~\cite{spes}, 67/232 for
EQUITAS~\cite{equitas} under set semantics, and 34 of 39 supported
cases for UDP~\cite{udp}. 
Although each pair is associated with a Calcite rule, the suite is
not a pure rule-invocation benchmark: The 22 missed cases also depend on
Calcite's surrounding normalization and decorrelation machinery or the exact rule-registration configuration,
so registering the rules alone does not guarantee a proof.

The LLM-generated rewrites are the most difficult class: \sys{}
proves 97 of 240 expected-equivalent pairs (\textbf{40.4\%}).
Coverage differs substantially by rewriter, reaching 64/127
(\textbf{50.4\%}) for SPA-32B-SI and 33/113
(\textbf{29.2\%}) for GPT-5.4. SPA-32B-SI more often produces
CTE-and-join forms, whereas GPT-5.4 more often introduces
\texttt{EXISTS} and \texttt{IN} structures that require additional
alignment. 


\subsection{Comparison with existing provers}\label{sec:xprover}

We compare \sys{} with QED~\cite{qed} and
SQLSolver~\cite{sqlsolver} on the 331 LLM-generated rewrite pairs
described in \S\ref{sec:workloads}.
We consider \emph{positive equivalence claims}: \texttt{PROVED} for \sys{}, \texttt{PROVABLE} for QED, and
\texttt{EQ} for SQLSolver.  
Table~\ref{tab:prover-comparison} shows that \sys{} proves 97 of
the 240 execution-result-identical pairs, compared with 35 for QED and 38 for SQLSolver. The union of the three systems proves 114 pairs,
17 more than \sys{} alone, showing that their supported reasoning
fragments are complementary.




\begin{table}[t]
  \centering
  \footnotesize
  \caption{Equivalence proof coverage on the 240
    expected equivalent LLM-generated rewrite pairs. }
  \label{tab:prover-comparison}
  \vspace{-3mm}
  \begin{tabular}{@{}lrrr@{}}
    \toprule
    & \sys{} & QED & SQLSolver \\
    \midrule
    Equivalence proofs
      & \textbf{97} & 35 & 38 \\
    Proof rate
      & \textbf{40.4\%} & 14.6\% & 15.8\% \\
    Claims unique to prover
      & \textbf{58} & 8 & 5 \\
    \bottomrule
  \end{tabular}

  \vspace{-3mm}
\end{table}

\noindent\textbf{Runtime.}
For successful proofs, the three systems have median runtimes of the
same order: 1.3\,s for \sys{}, 0.7\,s for QED, and 1.3\,s for
SQLSolver. For \sys{}, the 90th percentile is 2.3\,s and the maximum
is 3.8\,s; for QED, the corresponding values are 2.8\,s and 43\,s;
SQLSolver's maximum is 58\,s. Their unsuccessful runs differ more
substantially. \sys{} exhausts all configured search attempts in a
median of 11.4\,s. QED's undecided runs take a
median of 10.6\,s. For SQLSolver, 145 of 331 runs reach its approximately 60\,s timeout. 

\section{What Limits Rewrite-Based Proofs?}\label{sec:limits}





\subsection{Which transformations are most crucial?}
\label{sec:concentration}

The final configuration contains many transformations, but frequent
use does not imply necessity: when one transformation is removed,
the proof still holds, or another derivation may still prove the same pair. We therefore
distinguish \emph{usage}, whether a transformation appears in a
recorded proof, from \emph{leave-one-out necessity}, whether the proof
is lost when that transformation is disabled.

We first examine the transformations appearing in 764 recorded
proofs from BIRD, LeetCode, the literature suite, and the two
LLM-rewrite corpora. We then evaluate the twenty most frequently used
transformations by disabling each one in turn while leaving the
remaining configuration and search budgets unchanged. A pair is
\emph{dependent} on a transformation when it is proved in the full
configuration but not when that transformation is disabled.

The results show that necessity is highly concentrated. The
join-region-ordering canonicalizer appears in 67 proofs and is
individually necessary for 40 of them, a leave-one-out loss rate of
59.7\%. The loss counts for the other nineteen measured transformations
sum to 45, and no other individual transformation is necessary for
more than 14 proofs. Global alignment of commutative join regions is
therefore the dominant individually load-bearing component in this
analysis. Usage alone gives a very different picture. Ten of the twenty
most-used transformations are individually necessary for no proof.
Several join--project transpose rules appear in
42--46 proofs each, yet disabling any one of them loses at most one
proof.
Moreover, disabling all ten zero-loss transformations together preserves
every proof among the 112 pairs that use at least one of them. These
transformations may participate in successful derivations, but the
remaining rule library usually provides an alternative route to the
same witness.




\subsection{Remaining gaps}
\label{sec:gaps}

We classify every execution result identical pair that remains unproved on the two LLM-rewrite suites. The analysis also considers SQL-level query features.
Table~\ref{tab:gap-mechanics} reports the terminal search condition
for each pair. 

\begin{table}[t]
  \centering
  \footnotesize
  \caption{Terminal conditions for unresolved execution-identical
    pairs in the LLM-rewrite suites.}
    \vspace{-3mm}
  \label{tab:gap-mechanics}
  \begin{tabular}{@{}lrrr@{}}
    \toprule
    Terminal condition & SPA-32B-SI & GPT-5.4 & Total \\
    \midrule
    Rule-application cap reached & 40 & 47 & 87 \\
    Search saturated without overlap & 14 & 25 & 39 \\
    Timeout & 7 & 7 & 14 \\
    Input load failure & 2 & 1 & 3 \\
    \midrule
    \textbf{Total} & \textbf{63} & \textbf{80} & \textbf{143} \\
    \bottomrule
  \end{tabular}
  \vspace{-3mm}
\end{table}

The terminal conditions distinguish two principal limitations. In
101 cases, search reaches either the rule-application cap (number of total rules applied) or the timeout, often while exploring large join regions. These cases
suggest a search-capacity or pruning problem: the exploration has not
saturated when it is stopped. In another 39 cases, the memo saturates
after relatively few transformations without connecting the two
plan memos. These failures provide stronger evidence of a missing
transformation or canonical representation. 

Wide-join CTE structures dominate the unresolved cases. Queries
joining at least five tables account for 76\% of the SPA-32B-SI gaps
and 76\% of the GPT-5.4 gaps, compared with 62\% and 48\% of their
proved pairs. Their large join regions both increase search cost and
make complete structural alignment more difficult.
There are several more specific patterns. In SPA-32B-SI,
these include reordered \texttt{UNION} branches and existence tests
moved across CTE boundaries or re-keyed through joins. In GPT-5.4,
outer-join expressions and window-function rewrites are prominent;
the current transformation library contains no general window
normalization. Differences between scalar aggregation and
materialized aggregate CTEs also remain common in both suites.

\section{Conclusion}\label{sec:conc}

Recent work has shown that optimizer transformations can be repurposed to verify query equivalence. We ask what it takes to turn such query-rewriting knowledge into a standalone SQL equivalence prover. Existing rewrite rules already cover a substantial fraction of the equivalences in our workloads: \sys{} proves 974 of 1,612 expected-equivalent pairs across three workload classes, including 97 of 240 expected-equivalent LLM rewrites, compared with 38 for SQLSolver and 35 for QED.
Our experiments show that proof coverage depends less on larger search budgets than on two capabilities: aligning equivalent queries to common representatives and applying transformations under schema constraints. By recording the transformations, their applicability conditions, and the resulting common witness, \sys{} turns query-rewriting knowledge into inspectable and replayable equivalence proofs beyond black-box EQ judgments.

\bibliographystyle{ACM-Reference-Format}
\bibliography{main}

\end{document}